\documentclass{elsart}

\usepackage{natbib}

\usepackage{graphics}
\usepackage{epsfig}

\usepackage{amssymb}

\begin{document}

\begin{frontmatter}



\title{Interest Rates Mapping}


\author[IGAR]{M.Kanevski\corauthref{cor}},
\ead{Mikhail.Kanevski@unil.ch}
\author[BCGE]{M.Maignan},
\author[IGAR]{A.Pozdnoukhov\thanksref{SNSF}},
\author[IGAR]{V.Timonin\thanksref{SNSF}}

\thanks[SNSF]{Supported by the Swiss National Science Foundation projects ``GeoKernels'' 200021-113944 and ``Clusterville'' 100012-113506.}

\address[IGAR]{Institute of Geomatics and Analysis of Risk (IGAR), Amphipole building, University of Lausanne, CH 1015 Lausanne, Switzerland}

\address[BCGE]{Banque Cantonale de Gen\`{e}ve (BCGE), Geneva, Switzerland}

\corauth[cor]{Corresponding author.}

\begin{abstract}
The present study deals with the analysis and mapping of Swiss franc interest rates. Interest rates depend on time and maturity, defining term structure of the interest rate curves (IRC). In the present study IRC are considered in a two-dimensional feature space - time and maturity. Geostatistical models and machine learning algorithms (multilayer perceptron and Support Vector Machines) were applied to produce interest rate maps. IR maps can be used for the visualisation and patterns perception purposes, to develop and to explore economical hypotheses, to produce dynamic asses-liability simulations and for the financial risk assessments. The feasibility of an application of interest rates mapping approach for the IRC forecasting is considered as well.
\end{abstract}

\end{frontmatter}

\section{Introduction}
The present study deals with an empirical analysis and mapping of Swiss franc (CHF) interest rates (IR). The complete empirical analysis includes comprehensive quantitative analysis and characterisation of daily behaviour of CHF interest rates from October 1998 to December 2005. A mapping part of the paper considers the application of spatial interpolation models for IR mapping in a feature space ``maturity-date''. In a more general setting interest rates can be considered as functional data (IR curves are formed by different maturities) having specific internal structures (term structure) and parameterised by econometric models.

The main issues related to IR mapping are the following: 1) empirical analysis of IR spatio-temporal patterns, 2) reconstruction and prediction of interest rate curves; 3) incorporation of economical/financial hypotheses into the IR prediction process; 4) development of what-if scenario for financial engineering and risk management. Some of the preliminary ideas elaborated in this study  were firstly presented in \citep{ref1}.

In general there are two principal approaches to make term-structure predictions \citep{ref2}: no-arbitrage models and 2) equilibrium models. The no-arbitrage models focuses on fitting the term structure at a point in time (one dimensional model depending on maturity) to ensure that no arbitrage possibilities exist. This is important for pricing derivatives. The equilibrium models focuses on modelling the dynamics of the intravenous rate using affine models after which rates at other maturities can be derived under various assumptions about risk premium. Detailed discussion along with corresponding references can be found in \citep{ref2}.

An important and interesting approach complementary to classical empirical analysis of interest rates time series was developed in \citep{ref3, ref4, ref5} where both traditional econophysics studies (power law distributions, etc.) and a coherent hierarchical structure of interest rates were considered in detail. An empirical quantitative analysis of multivariate interest rates time series and their increments (carried out but not presented in this paper) includes study of autocorrelations, cross-correlations, detrending fluctuation analysis, embedding, analysis of distribution of tails, etc. \citep{ref1, ref5, ref6, ref7}.

The most important part of the current study deals with an IR mapping in a two dimensional feature space \{maturity(months), time(date/days)\} using spatial interpolation/extrapolation models (inverse distance weighting - IDW, geostatistical kriging models), nonlinear artificial neural network models (multilayer perceptron - MLP) and robust approaches based on recent developments in Statistical Learning Theory (Support Vector Regression) \citep{ref10}. Embedding of IR data into a two-dimensional space brings us to the application of spatial statistics and its modelling tools. Higher dimensional feature spaces can be considered and applied as well. Simple models, like linear and inverse distance weighting were used mainly for the comparison and visualisation purposes.

\begin{figure}
\centering
\includegraphics[width=6.5cm]{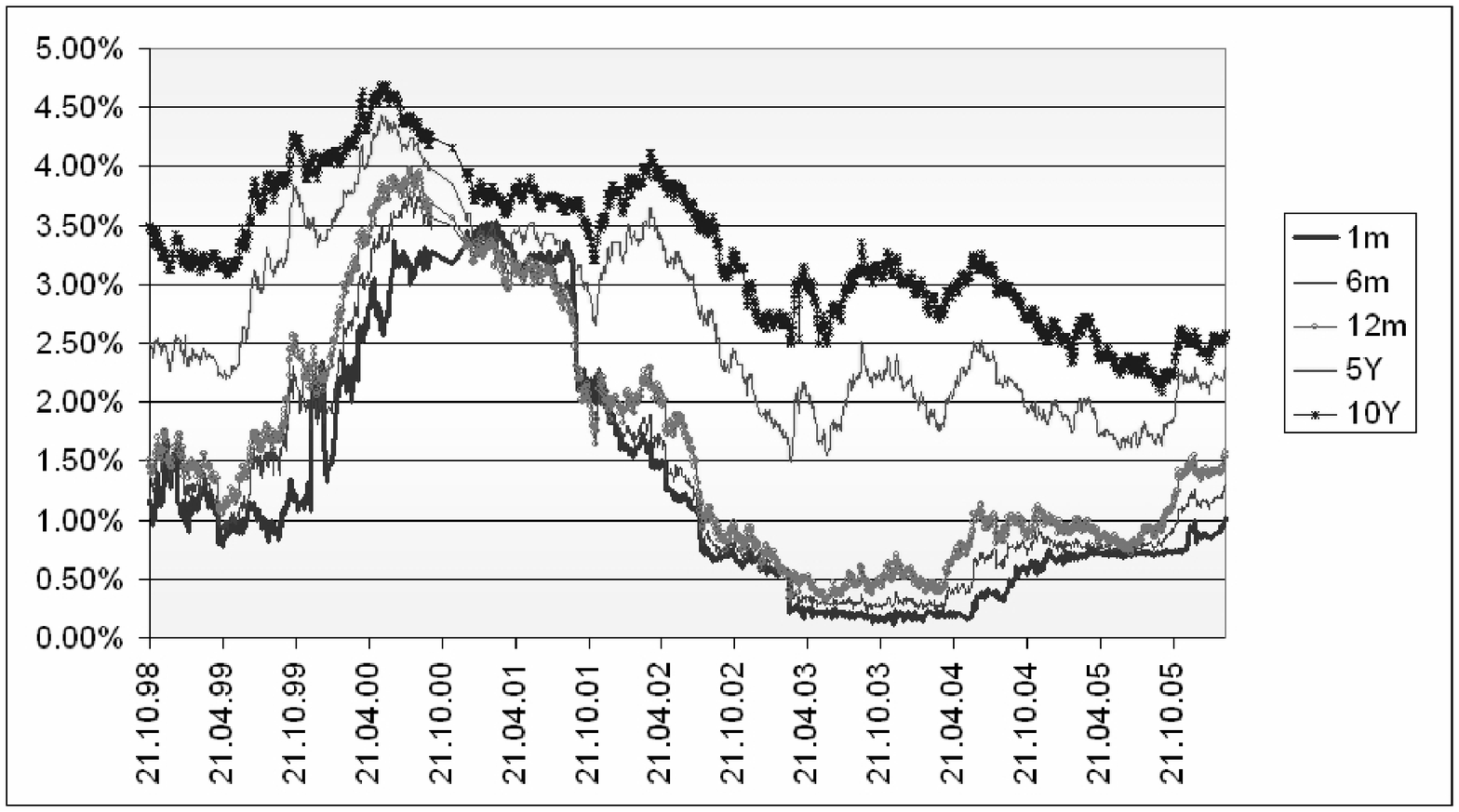}
\includegraphics[width=6.5cm]{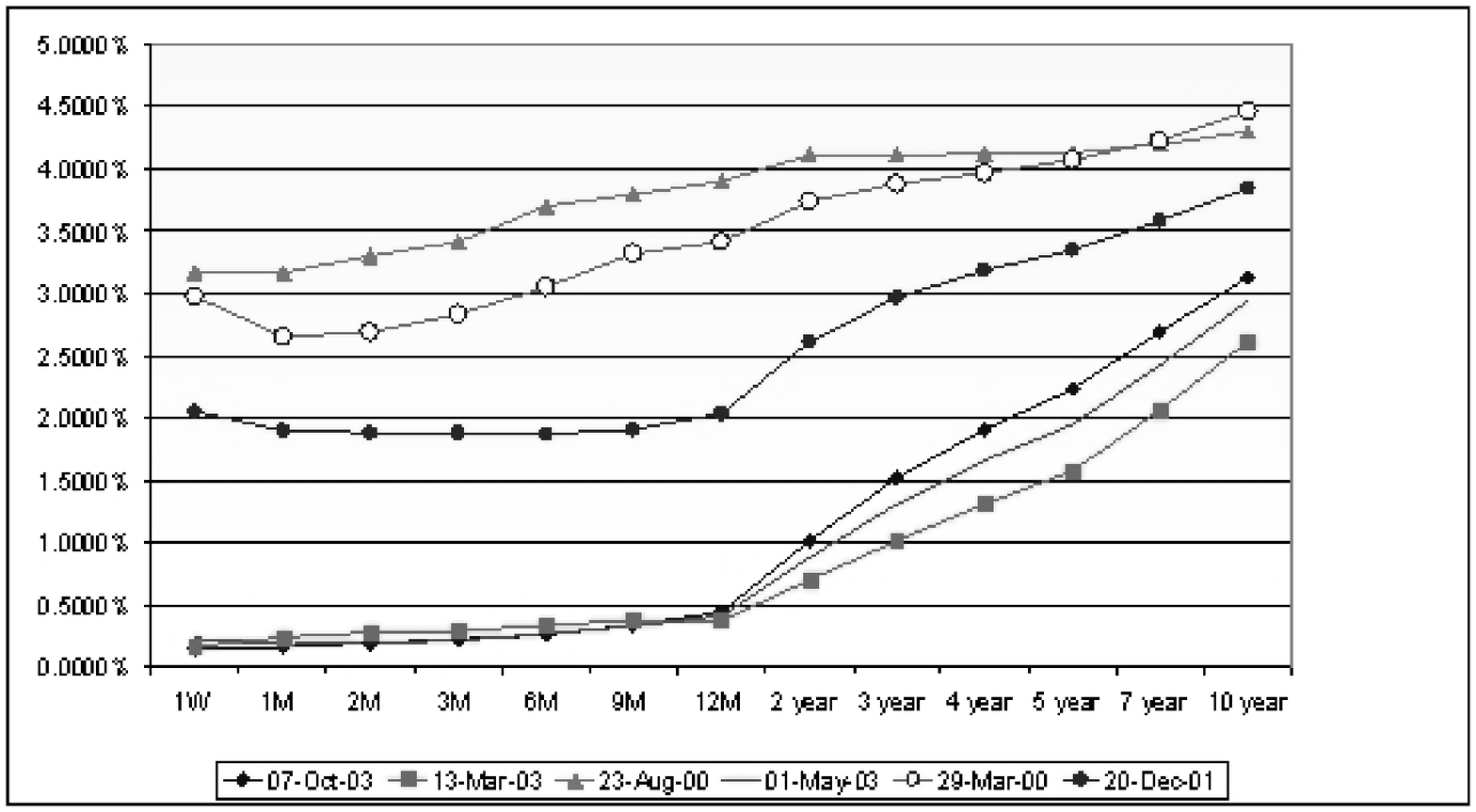}
\caption{Left: Evolution of IR time series for different maturities: 1,6,12 months and 5 and 10 years. Right: Examples of the observed CHF
interest rates curves for different days.}
\label{fig1}
\end{figure}

Evolution of CHF interest rates data is given in Figure~1 where temporal behaviour of different maturities is presented. The IRCs are composed of LIBOR interest rates (maturities up to 1 year) and of swap interest rates (maturities from one year to 10 years). Such information is available on the specialised terminals like Reuters, Bloomberg, etc. and is usually provided for some fixed time intervals (daily, weekly, monthly) and for some definite maturities (in this research we use the following maturities: 1 week, 1, 2, 3, 6 and 9 months; 1, 2, 3, 4, 5, 7 and 10 years).

There are some important stylized facts that have to be considered when modelling IR curves \citep{ref2}: the average yield curve is increasing and concave; the yield curve assumes a variety of shapes through time, including upward sloping, downward sloping, humped, and inverted humped; IR dynamics is persistent, and spread dynamics is much less persistent; the short end of curve is more volatile than the long end; long rates are more persistent than short rates. There is a coherence in evolution of interest rate curves (see some typical examples in Figure~1).

In the present research the application of IR mapping is concentrated on: 1) visualisation and perception of complex data: IR data represented as maps are easier for the analysis and interpretation, 2) reconstruction of interest rates at any time and for any maturities by using spatial prediction models and 3) prediction of IR curves. The last problem can be considered from two different points of view: a) predictions without any a priori information (problem of extrapolation using historical IR data) and b) forecasting under some prior hypotheses about the future evolution of market or interest rates (some banks provide information about future level of interest rates at several maturities). In the latter case one can consider different scenario ``what-if'' and prepare information on IR curves for dynamic simulations of portfolio development, asset-liability management etc. \citep{ref1}.

\section{Data description and interest rates mapping}
Daily CHF IR data from October 1998 to December 2005 were prepared as a training data set (data used to develop a model). The study of this period is quite interesting because during this time different market states (bullish, bearish) and even historically the lowest LIBOR rates were observed. Training data were used for IR mapping (IRC reconstruction and in-sample predictions) and to develop a prediction model which then was used to make predictions for January 2006 (out-of-sample predictions). Thus, real data for 2006 were used only for validation purposes.

Time series for several maturities are presented in Figure~\ref{fig1} (left). It is evident that the behaviour of the curves is rather coherent but very complex in time reflecting financial market evolution. Figure~\ref{fig1} (right) demonstrates term structure of several IR curves: dependence of IR on maturity for fixed dates. In December 21 an inversion of the curve, when very short interest rates (weekly) are higher than monthly interest rates was observed.

Despite of some coherencies in temporal behavior between different maturities, the relationship between short (1 month) and long (10 years) maturities are nontrivial which can be demonstrated by analysing correlation matrix. Sometimes multi-valued relationships can be detected.

In \citep{ref2} monthly IR were modelled using parametric Nelson-Siegel model based on 3 factors corresponding to long-term, short-term and medium term IR behaviour. These parameters can be interpreted in terms of level, slope ($\textrm{maturity}_{10years} - \textrm{maturity}_{3months}$) and curvature ($2\textrm{maturity}_{2years} - \textrm{maturity}_{3months} - \textrm{maturity}_{10years}$). First, these parameters were modelled as a time series using historical data and then they were applied to make IR forecasting. Such approach seems to be quite interesting and transparent both from scientific and practical point of views. The results obtained using linear time series models have demonstrated efficiency of this methodology. Possible extensions to nonlinear models were not considered and are a topic of ongoing research.

Let us consider interest rates mapping in a two-dimensional feature space described by maturity (X-axes, in months) and time (Y-axis, in days). In this presentation along X-axis one can observe IR curves for fixed date (when Y is fixed) and along Y-axis one can observe temporal behaviour of IR for fixed maturities. Distance between points in this space is rather synthetic and should be taken into account both during model development and interpretation of the results.

Geostatistical approach (family of kriging models) is based on empirical analysis and modelling of variograms \citep{ref9}. Variography was efficiently used to quantify the quality of machine learning algorithms modelling by estimating the spatial structure of the MLA residuals: good models have to demonstrate pure nugget effects (no spatial correlations) with a variance fluctuating around a raw data noise level \citep{ref9}.

Let us consider interest rate mapping procedure. The value at each prediction (unsampled) point $Z(x,y)$ can, in general, be assessed in two ways: 1) model $Z_{1}$ is a weighted sum of the measured/observed neighbouring data $Z_{i}$ or 2) model $Z_{2}$ is a weighted sum of kernel functions $K_{j}$:
\[
\begin{array}{l}
 Z_{1} (x,y) = {\sum\limits_{i = 1}^{n} {w_{i}(x,y) Z_{i} (x_{i} ,y_{i} )}} {\rm}     \\
 Z_{2} (x,y) = {\sum\limits_{j = 1}^{m} {\alpha _{j} K_{j} (x,y;x_{i} , y_{i})}} + b
 \end{array}
\]
where $n$ and $m$ are the number of points or kernels used for the prediction, $w_{i}$ and \textit{$\alpha$}$_{i}$ are corresponding weight coefficients.

In case of inverse distance weighting mapping the weights are inversely proportional to the power of distance between observations and prediction point; in case of kriging models they are a solution of a system of linear equations derived from the principle of best linear unbiased predictor; and in case of Support Vector Regression they are a solution of quadratic optimization problem following Statistical Learning Theory \citep{ref8, ref9, ref10}. In case of MLP kernels are replaced by a combination of nonlinear transfer function, e.g. sigmoid. IDW, kriging and MLP are well known and widely applied models for spatial data analysis and mapping.

Statistical Learning Theory is a general framework for solving classification, regression and probability density estimation problems using finite number of empirical data. SVR is a non-parametric regression method, which exploits kernel expansion. It attempts at minimizing the empirical risk (the residuals on the training data), simultaneously keeping low the complexity of the model. By doing this, the over-fitting on the training data can be avoided and one may expect promising predictive abilities. Unlike MLP, SVR after fixing few hyper-parameters, has a unique solution of quadratic optimisation problem \citep{ref10}.

All models (IDW, kriging, MLP, SVR) depend on some hyper-parameters (number of neighbours, IDW power, variogram model, number of hidden neurons, regularization and kernel parameters, etc.) which can be tuned using different statistical techniques, e.g. data splitting (training-testing-validation), cross-validation, jack-knife.

\begin{figure}

\centering
\includegraphics[width=6.5cm]{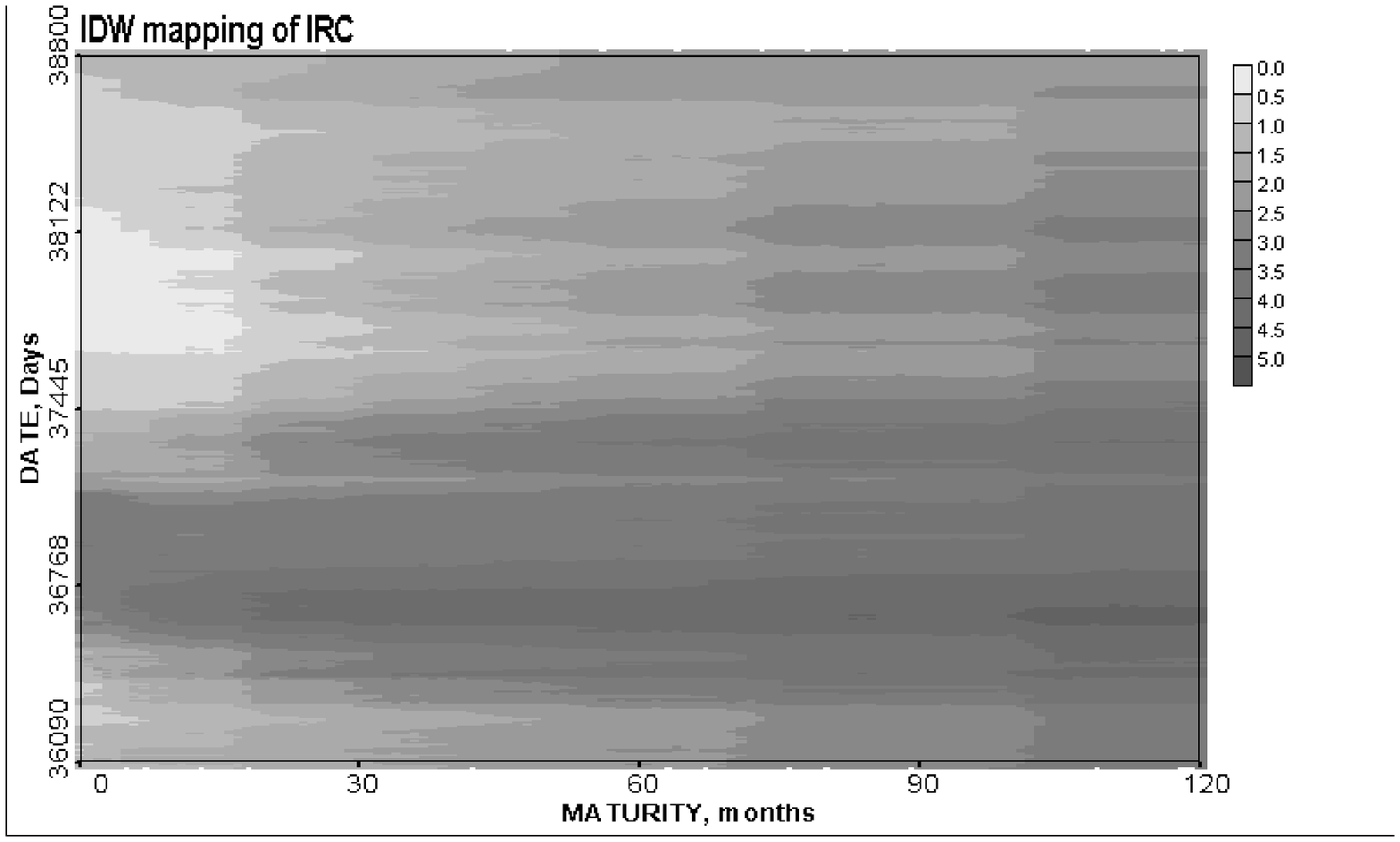}
\includegraphics[width=6.5cm]{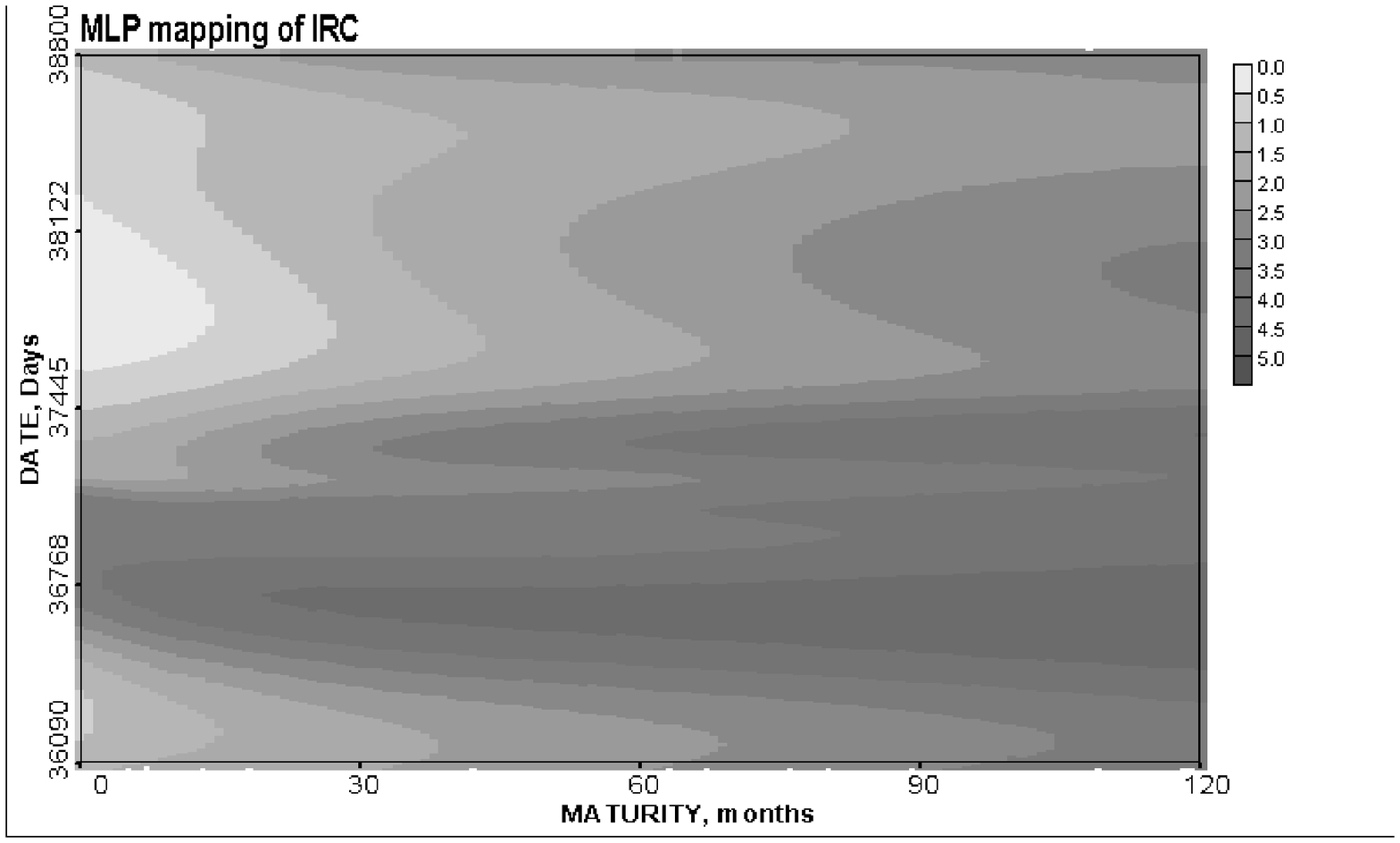}
\caption{Interest rates mapping using inverse distance weighting (left) and multilayer perceptron model (right).}
\label{fig3}
\end{figure}

The result of IDW mapping with a power equal to two is presented in Figure~\ref{fig3}. This map gives quite consistent reproduction of the evolution of interest rates in time. IDW model is fast to tune and easy to apply and useful for the visualisation purposes. But it does not take into account spatial structure of data and can produce some artifacts.

Second approach presented in this paper is based on MLP. It is a well known nonlinear modelling tool having both advantages and some drawbacks as a black-box tool (interpretability of the results, overfitting of data, multiple local minima, etc.). Nevertheless, MLP is a flexible and very powerful approach for data mining and data modelling. The efficiency and quality of mapping of MLP can be controlled using (geo)statistical tools both for raw data and for the residuals \citep{ref8, ref9}. More details and a generic methodology of MLP application for spatial data analysis and mapping can be found, for example, in \citep{ref9} and for different financial applications in \citep{ref11}. The MLP map is given in Figure~\ref{fig3}.

Following the data modeling traditions we have split original IR data into training (80\% of data) and testing subsets (20\% of data). Testing data set was used to control the quality of different MLP and to avoid overfitting. Testing data set was used to find optimal number of hidden layers and hidden neurons. As an optimal a network with 2 hidden layers each consisting of 25 hidden neurons was applied. By comparing the results of MLP mapping and visualisation of original data it can be noted that MLP was able to detect and to model large and medium scale structures using nonlinear smoothing approach. Small scale and low level variability were ignored and treated as insignificant noise.

\begin{figure}
\centering

\includegraphics[width=6cm]{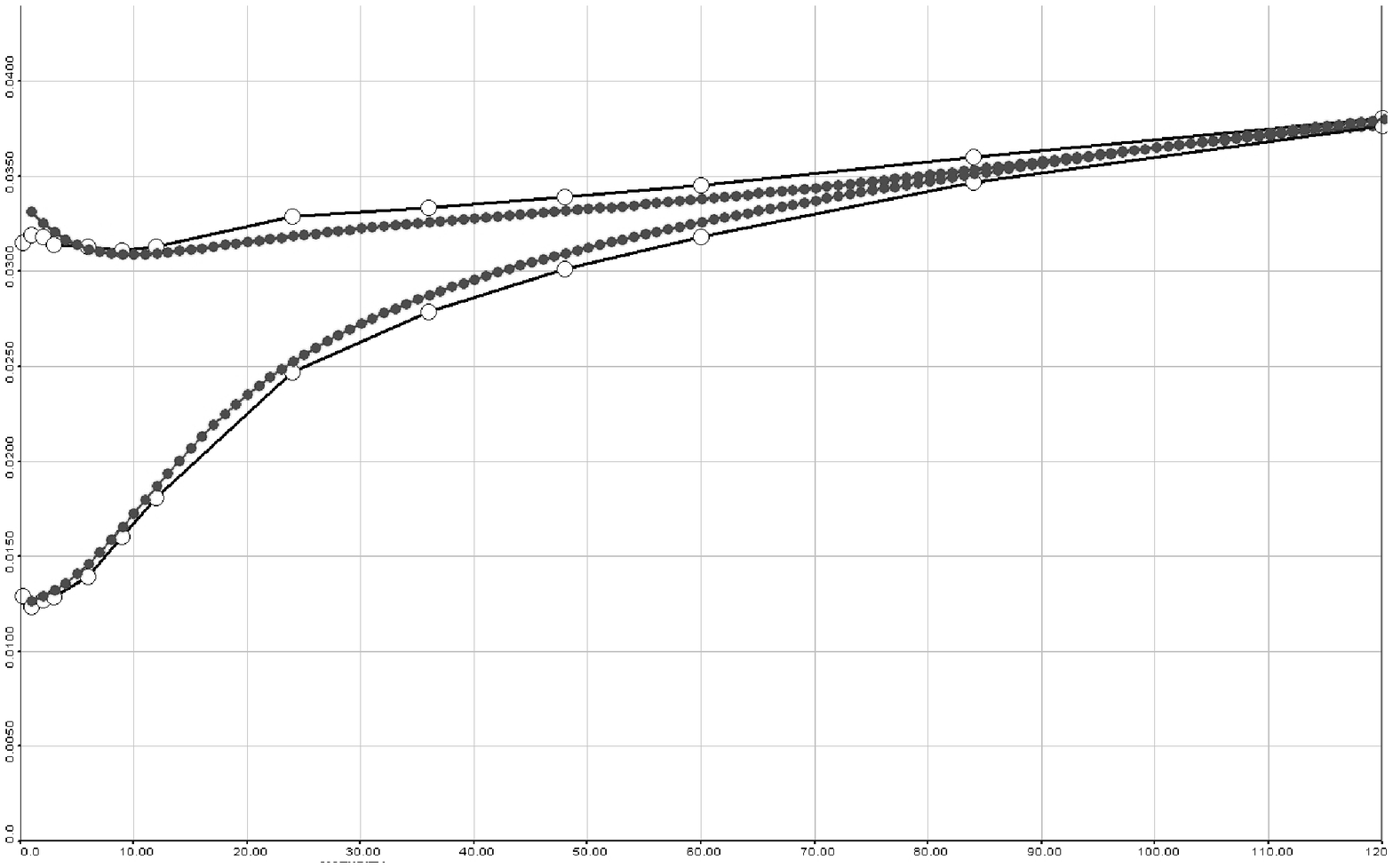}
\includegraphics[width=7.4cm]{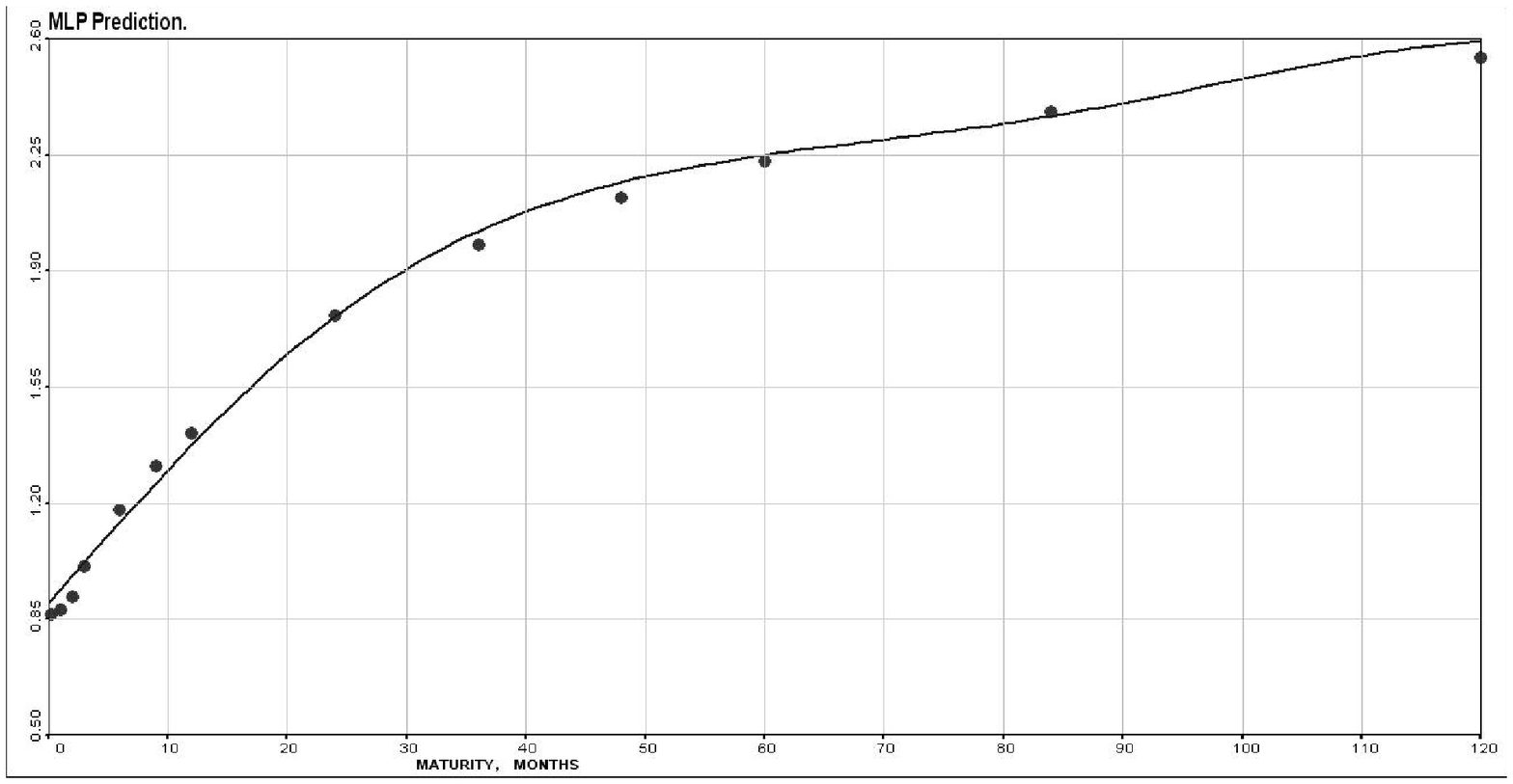}
\caption{Left: CHF interest rate curves reconstruction using MLP. Right: CHF interest rate curve prediction using MLP.}
\label{fig5}
\end{figure}

Reconstruction of two different interest rate curves (in-sample prediction) - these curves were extracted from training data set and then modelled using MLP is given in Figure~\ref{fig5}. Both curves are quite well reconstructed, including inversion of IRC.

Developed MLP model was used for the one month ahead IRC prediction. An example of the predicted IRC along with real/observed data is given in Figure~\ref{fig5}. The result is quite promising and more simulations under different market conditions and time horizons are in progress to validate the predictability of this mapping approach.

\begin{figure}

\centering
\includegraphics[width=6cm]{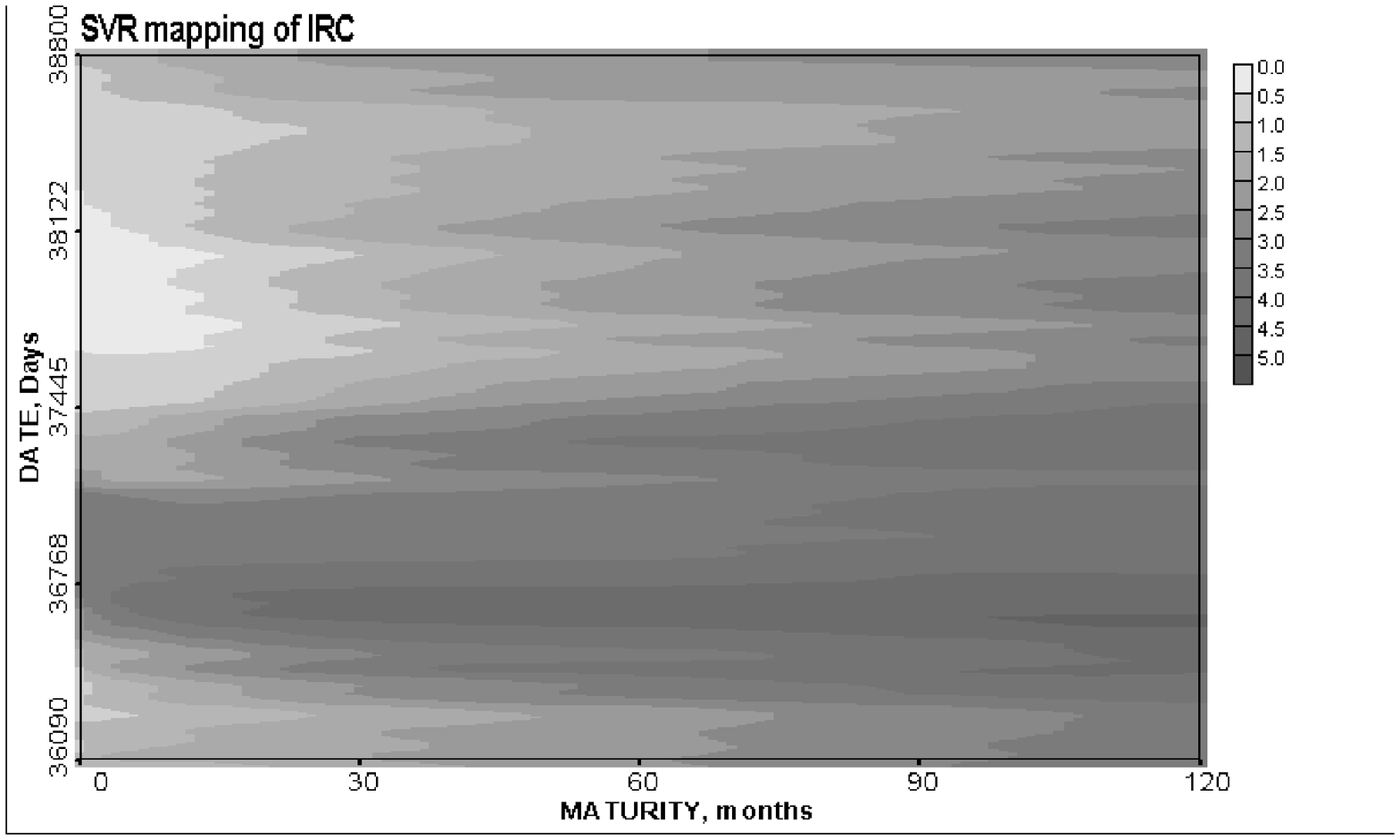}
\includegraphics[width=7cm]{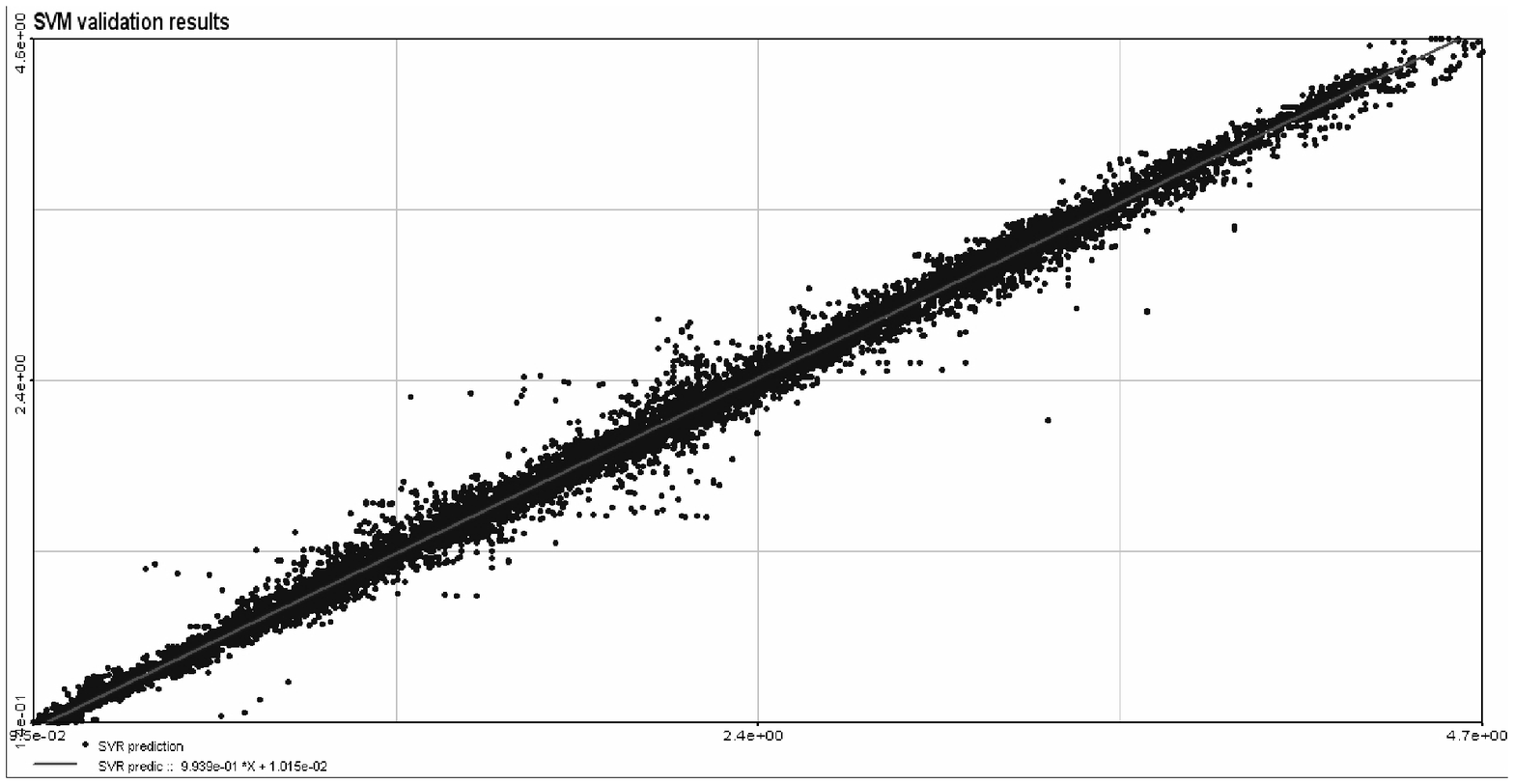}
\caption{Left: Interest rates mapping using Support Vector Regression. Right: Testing of Support Vector Regression model.}
\label{fig7}
\end{figure}

Finally, Support Vector Regression was applied for CHF IR mapping. The same as for MLP methodology was applied to tune SVR hyper-parameters: epsilon-insensitive zone of robust loss-function, regularisation parameter C, and the shape of a Gaussian kernel used. It should be noted that SVR approach is quite computationally intensive and tuning of its parameters is not a trivial task. The IR map produced by SVR is presented in Figure~\ref{fig7}. In comparison with MLP SVR is less smoothing data and well reproduces spatio-temporal pattern.

In order to validate the quality of SVR the testing data predictions (also not used to develop a model) are compared with the observed ones (Figure~\ref{fig7}, right). The results obtained are quite promising.
An important improvement for all models proposed can be achieved when applying the same analysis (development and retraining of models) within a moving window which can be related to the market conditions. This can partly help to avoid also the problem of spatial nonstationarity. Another approach can be based on hybrid MLA+geostatistics models \citep{ref9} already successfully applied for spatial environmental data. All spatial data analysis was carried out using Geostat Office software \citep{ref9}.

\section{Conclusions}
\label{Conclusions}
The paper presents the results on mapping of CHF interest rates. An important attention was paid to the visualisation of IR curves in a two-dimensional maturity-date feature space. Tools and models from spatial statistics and machine learning were applied to produce the maps. Developed interest rates maps can be easily interpreted and give good summary view on IRC evolution. Some promising results on forecasting were obtained using artificial neural networks. An important improvement can be achieved by using hybrid models based on geostatistics and machine learning and incorporating time series tools into modelling and forecasting procedures. In the future IR mapping methodology will incorporate economical/financial hypotheses of market behaviour in order to elaborate ``what-if'' scenarios for the financial risk management.





\begin{thebibliography}{}


\bibitem[Kanevski et. al.(2003)]{ref1} Kanevski M., Maignan M., and Michoud D. (2003). Spatial Statistics for Finance: Interest Rates Mapping. Presented at Swiss Statistical Society conference. Montreux, Switzerland.

\bibitem[Diebold and Canlin(2006)]{ref2} Diebold F. and Canlin Li (2006). Forecasting the term structure of government bond yields. Journal of Econometrics. 130:337-364.

\bibitem[Di Matteo et. al.(2005)]{ref3} Di Matteo T., Aste T., Hyde S.T. and Ramsden S. (2005). Interest rates hierarchical structure. Physica A 355:21-33.

\bibitem[Di Matteo and Aste(2002)]{ref4} Di Matteo T., Aste T.  (2002). How does Eurodollars interest rates behave? Journal of Theoretical and Applied Finance 5:122-127.

\bibitem[Cajueiro and Tabak(2007)]{ref5} Cajueiro D. and Tabak B. (2007). Long-range dependence and multifractality in the term structure of LIBOR interest rates. Physica A, v.373, pp.603-614.

\bibitem[Mantegna and Stanley(1999)]{ref6} Mantegna R. and Stanley H. (1999). An Introduction to Econophysics: Correlations
and Complexity in Finance. Cambridge Univ. Press, 158 p.

\bibitem[Kantz and Schreiber(2003)]{ref7} Kantz H. and Schreiber Th. (2003). Nonlinear Time Series Analysis.
Cambridge University Press, Cambridge.

\bibitem[Haykin(1999)]{ref8} Haykin S. (1999). Neural Networks. A Comprehensive Foundation. Prentice-Hall International. London, 842 p.

\bibitem[Kanevski and Maignan(2004)]{ref9} Kanevski M. and Maignan M. (2004). Analysis and Modelling of Spatial Environmental Data. EPFL Press, Lausanne, Switzerland.

\bibitem[Vapnik(1999)]{ref10} Vapnik V. (1999) The Nature of Statistical Learning Theory. 2$^{{\rm n}{\rm d}}$ edition. Springer, N.Y. 314p.

\bibitem[McNelis(2005)]{ref11} McNelis P.D. (2005). Neural Networks in Finance. Gaining Predictive Edge in the Market. Elsevier, Burlington, MA.



\end{thebibliography}
\end{document}